\documentclass[12pt]{article}

\usepackage{graphics}
\usepackage{amsmath}
\def\ni{\noindent}
\def\ph{{\phantom{...}}}
\def\={\phantom{..} = \phantom{..}}

\def\+{\phantom{..} + \phantom{..}}
\def\>{\phantom{..} > \phantom{..}}
\def\<{\phantom{..} < \phantom{..}}
\def\-{\phantom{..} - \phantom{..}}

\def\bq{\begin{quote}}
\def\eq{\end{quote}}
\def\be{\begin{equation}}
\def\ee{\end{equation}}
\def\bar{\begin{eqnarray}}
\def\ear{\end{eqnarray}}
\def\no{\nonumber}

\def\qp{quantum physics}

\def\prob{probability}
\def\probs{probabilities}
\def\ea{{\em et al. }}

\def\Sch{Schr{\"o}dinger}

\def\Schists{Schr{\"o}dingerists}

\def\Schs{Schr{\"o}dinger's}

\def\Copism{Copenhagenism}

\def\Copists{Copenhagenists}

\def\vN{von Neumann}

\def\Poin{Poincar{\'e}}
\def\Ham{Hamiltonian}
\def\wf{wavefunction}

\def\cH{{\bf{H}}}

\def\sjN{\sum_{j=1}^N}
\def\sijN{\sum_{i,j=1}^N}
\def\skN{\sum_{k=1}^N}

\def\srmo{{\sqrt{ - 1}}}
\def\half{{\frac{1}{2}}}

\title{\bf On Non-Linear Quantum Mechanics\\[1in] and the Measurement Problem\\[1in] IV. Experimental Tests\\[2in]}

\author{W. David Wick\footnote{email: wdavid.wick@gmail.com}}

\begin{document}
\maketitle
\pagebreak

\section*{Abstract}

I discuss three proposed experiments that could in principle locate the boundary between the classical and quantum worlds, 
as well as distinguish the Hamiltonian theory presented in the first paper of this series from
the spontaneous-collapse theories.

\section{Introduction}

Ninety years post-quantum-revolution, we still do not know where, or what, is the Infamous Boundary. 
(This phrase, coined by John Bell, \cite{bell}, was suggested by William Faris for the title of a book of ours that appeared in 1995, \cite{tib}.)
Of course, many suggestions have been made on this topic, but most do not offer specific predictions that could be tested in the laboratory. 
Here I contrast the theory of paper I in a series, \cite{wick}, with the 
spontaneous-collapse (SC) theories, each of which do make predictions, and discuss some proposed experiments.

The first SC theory was presented by G. C. Ghirardi, A. Rimini, and T. Weber in 1986, \cite{grw}. 
In their proposal, the fundamental paradox of quantum theory referred to as the Measurement Problem---namely, that in measurement scenarios
two or more wave packets representing different outcomes separate and move in different directions, with no implied result---was resolved by postulating a mechanism of random, spontaneous collapses.
These happen rapidly for a large (``macroscopic") system but rarely on the atomic level, thus preserving the successes of QM at that scale while explaining the adequacy of classical physics at the
larger scale. Variants have been proposed over the years, for instance continuous-collapse (CC), in which a continuous stochastic process substitutes for the jumps in the wavefunction, see e.g., \cite{cc}.
(CC replaces \Schs\ deterministic equation by a ``stochastic differential equation" of the ``Brownian motion plus drift" variety, but for the wavefunction rather than a particle.
The drift drives the wavefunction to lower spatial dispersion, while the ``Brownian" part somewhat opposes it. The magnitude of the random part is the square-root of the drift.
I am familiar with this kind of model from mathematical biology, where it is sometimes useful as a simplified, continuum description for a discrete population, \cite{war}. 
\Schists\ of course reject any interpretation of the wavefunction as an approximation to an underlying discrete system.)   

The collapse theories can be differentiated from the Hamiltonian theory of paper I by several characteristics. The latter (a) preserves energy exactly (trajectory-by-trajectory); (b) preserves the norm exactly;
and (c) is exactly time reversible. (In other words, it has all the familiar features of physicists' theories from decades and centuries past.) The former do not enjoy these properties. 
Both SC and CC theories have
at least two free parameters (governing rate and extent of collapses). The Hamiltonian theory has one (the coupling constant between the linear and nonlinear parts, denoted by ``$w$" in previous papers) and possibly another 
(the magnitude of a random part of the wavefunction, see paper II).
However, in paper III it was remarked that, as is typical of high-dimensional, nonlinear deterministic theories, the evolution may be chaotic. 
(But only an instability in measurement situations was demonstrated there.) If so, a parameterized model of a random part may be superfluous 
(for roulette, no one bothers to make a detailed model of the croupier's hand). 

There are claimed theorems asserting 
that any deterministic, nonlinear quantum theory must violate the no-action-at-a-distance rule of relativity (for a review see \cite{cc}), and therefore is unacceptable.
But, as for von Neumann's ``god is throwing dice" theorem of 1932, \cite{vN}, the argument is not decisive.\footnote{When 
Einstein was asked about the dice-playing god theorem, he pointed at von Neumann's Axiom II about summation of expected values and asked, ``Why should we believe in that"? 
Axiom II was unique to linear theories and ignored the impact of the measuring apparatus, see Bell's collected work, \cite{bell}, paper 1, or for the anecdote, \cite{wick}, p. 286.}
Assuming von Neumann's axioms for observables, which he derived from linear theory, at the outset renders the logic circular.  
\Schists\ need not agree that $\hbox{tr}\,\rho\, A$ for any self-adjoint operator $A$ represents an observable absent an explicit device description, nor that all observables are of that type.
(For example, in a nonlinear Hamiltonian theory the energy is not a von Neumann observable, as it is not quadratic in $\psi$.)
The coupled Dirac and Maxwell equations, with the Dirac charge current serving as the source terms for Maxwell's, 
provide an example of a relativistically-invariant, nonlinear quantum theory\footnote{It can be argued that Dirac's theory was not relativistically invariant, 
due to the nonunitary matrices implementing Lorenz transformations. 
The cure may require introducing an indefinite-metric Hilbert space, which spoils the Copenhagenist statistical interpretation of the wavefunction 
but \Schists\ might not object. See paper I and references therein.},  see \cite{barut}. 
For a suggested example in the present context, see paper I, section 4.

Concerning the IB: for collapse models, it is purely a function of scale, while for a Hamiltonian model it is a function of both scale and energy. 
That provides us with the experimental desideratum making distinguishing tests possible.

\section{An ``ideal" experiment}

\begin{figure}
\rotatebox{0}{\resizebox{5in}{5in}{\includegraphics{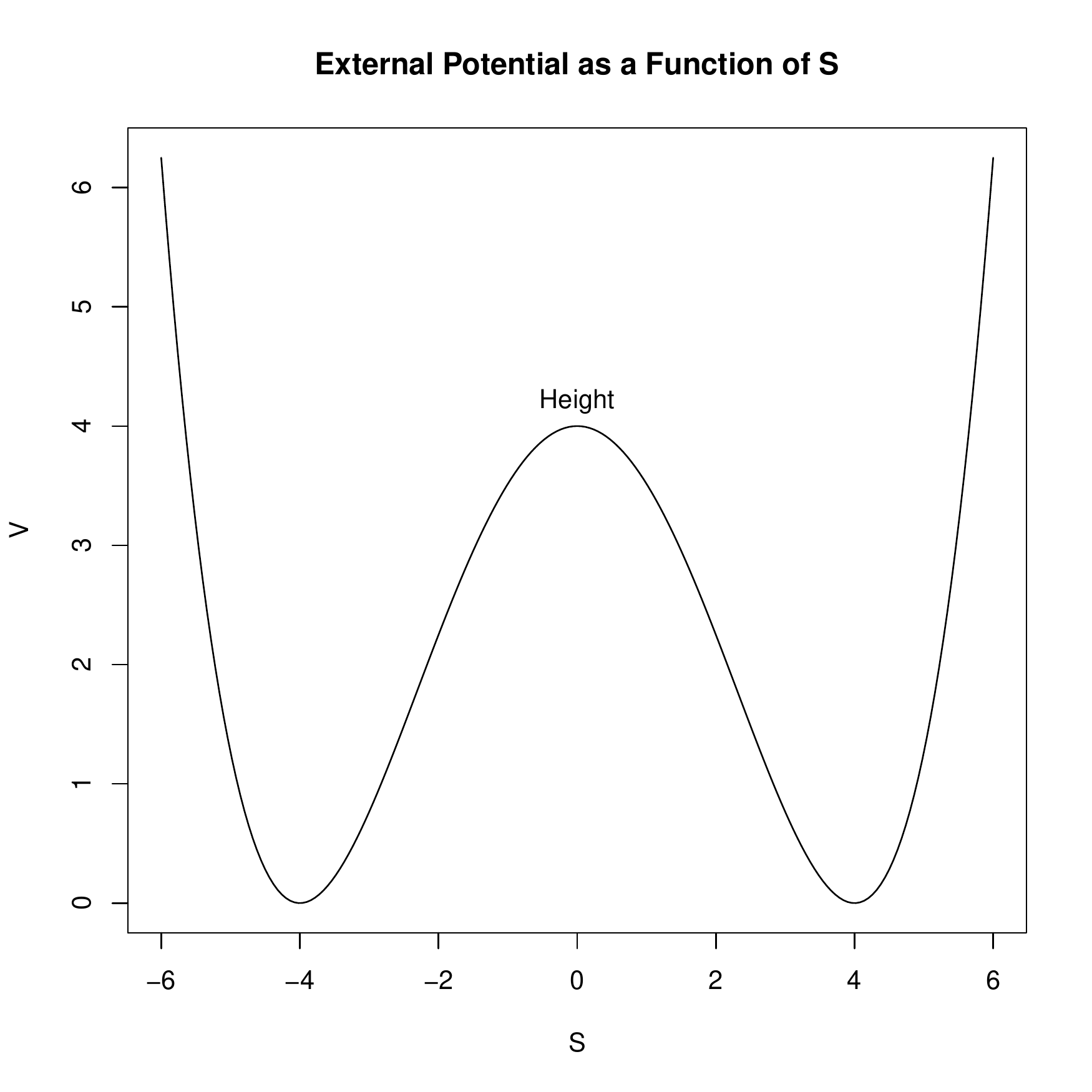}}}
\caption{``Double-well" external potential plotted vs. some ``spatial" degree of freedom.}\label{vfig}
\end{figure}

Consider a system of size conjectured to be near the IB, and subject to an external potential of the ``double-well" shape, see Figure \ref{vfig} (reproduced from paper III).
Initially the object should be confined (in dispersion) to a narrow band centered at the location of the central, unstable-equilibrium point (``hill") in the potential. 
In order to generate cats (if on the quantum side of the boundary), we can utilize one of two scenarios:

\begin{quote}
(a) Couple the ``macrosystem" to a microsystem, say one ``spin" (``qubit"), initially in a superposition of, say, spin-up and spin-down with equal weights. 
 In linear QM, wave packets describing the entangled state should separate, with the interpretations ``spin up and needle went up" and with down replacing up.

(b) Cool the system down to its groundstate. If it can be considered like a single quantum particle, it should oscillate in both wells simultaneously.
\end{quote}

\begin{figure}
\rotatebox{0}{\resizebox{5in}{5in}{\includegraphics{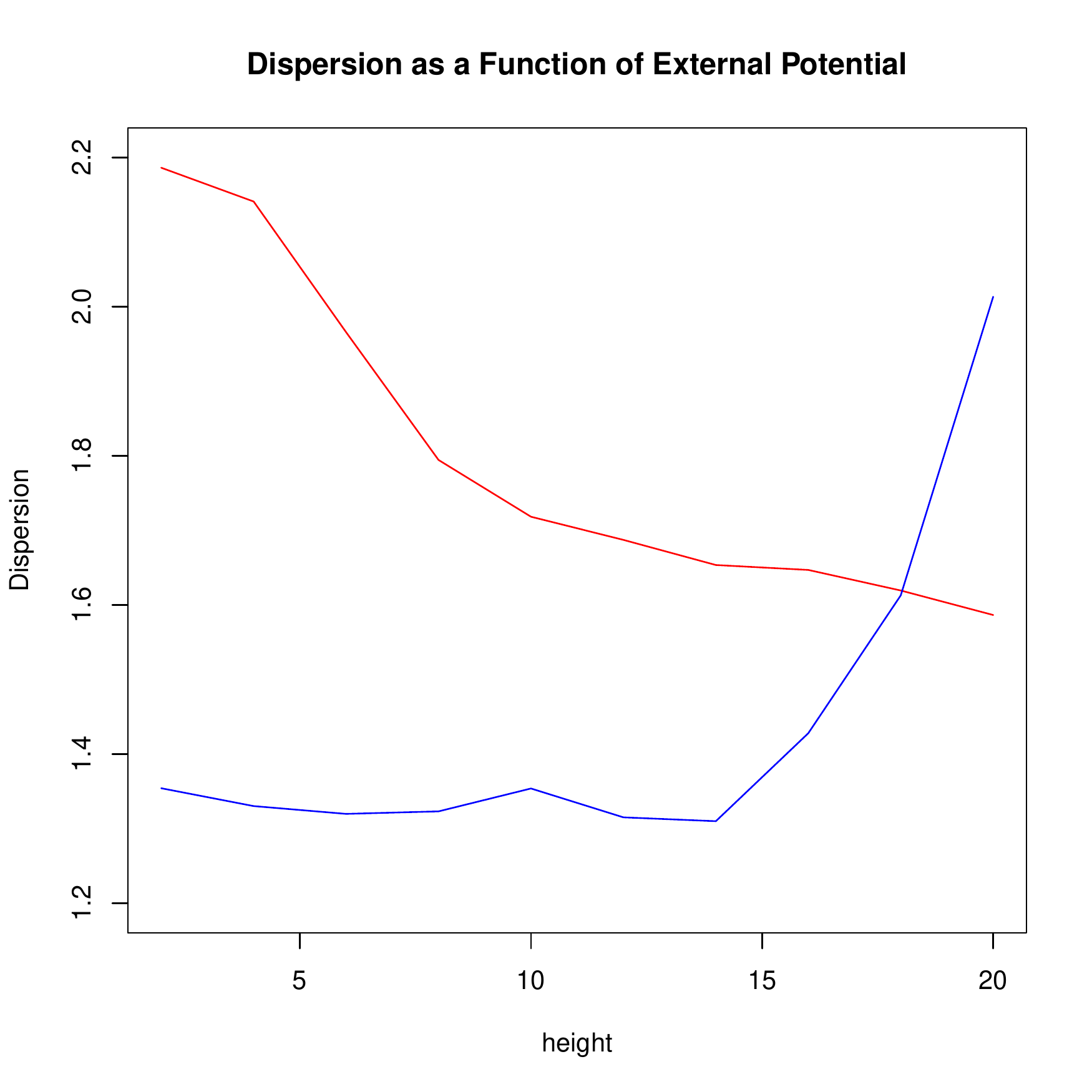}}}
\caption{Dispersion vs. height for the linear (red curve) and nonlinear (blue curve) models.}\label{dispfig}
\end{figure}

The goal is to observe cats or the absence thereof. 
The modern definition of ``cat" is a macroscopic (or at least directly observable) system whose dispersion is larger than its physical size. (See paper I for how
these quantities are defined by a wavefunction.)
Figure \ref{dispfig}, derived using the small-scale (``9 qubit") model from paper III, shows the dispersion as a function of the hill-height 
for the linear (w=0) and nonlinear (w=2.2) cases. The important element to note in this figure 
is the elbow or ``hockey stick" shape in the nonlinear case. This results from the energy barrier to forming cats: below the threshold the external potential cannot supply the required energy, 
while cats become possible above it. 
(It may be surprising---it was to this author---that in the linear case the dispersion decreases with the height parameter. 
For the explanation, see the Computational Appendix.)


Observing the hockey-stick could distinguish the Hamiltonian theory presented in paper I 
from the spontaneous-collapse theories. If the IB is purely a matter of scale,
when the system-size is below the IB, cats can form and perisist, but above a collapse quickly occurs, eliminating one of the cats. 
Thus we would expect to see two curves resembling the red curve in Figure 
\ref{dispfig}, one higher than the other (although perhaps the curves would display the reversed trends).

\section{Some proposed experiments}

Here are three criteria for an experiment that might both locate the IB and distinguish collapse theories from Hamiltonian:

\begin{quote}

(1) The system should be scalable.

(2) The system should be subjected to an adjustable external potential.

(3) The readout should make an unambiguous differentiation between cat-or-no-cat states.

\end{quote}

These are worth keeping in mind while reading the abbreviated accounts of some plausible experiments, presented next.

\subsection{The Marshall \ea experiment}

In the 1980s, L. Di{\'o}si, \cite{diosi}, and R. Penrose, \cite{penrosebook}, independently proposed a theory of collapse of the wavefunction due to gravitation. 
Roughly, the idea seems to be the following. 
Consider separating ``two lumps" (copies of some small object) by some distance.
This will require a certain cost in (gravitational) energy. In a time equal to Planck's constant divided by this potential energy, one packet is eliminated, due to ``uncertainty" 
(a concept that \Schists\ of course reject). This is explained as deriving from a putative ``quantum gravity theory" combined with another postulated ``gravitization of quantum mechanics", \cite{penrose}. 

Whatever the justification, this seems to be another spontaneous collapse theory, albeit one that does not assume a free parameter for the time before collapse. 
In 2002 Marshall and colleagues, \cite{marshall}, described an experiment to test the theory: a single photon scattered off a tiny mirror (size: one micron) suspended on a cantilever. 
Starting with a photon-present-or-photon-absent superposition state 
will entangle the mirror, with the latter's position dispersed
``by about the diameter of an atomic nucleus". Readout would be by fringes observed in a double-arm interferometer, from which a cat-state of the mirror would be inferred 
if interference reappeared with a multiple of the cantilever oscillation frequency. 

Mirrors are certainly scalable. 
But the predicted displacement is far smaller than the system size, making the cat appellation questionable. 
(Indeed, one could imagine that everything from gnats to planets is naturally dispersed by a femtometer---without calling into question the validity of classical mechanics.) 
There is no built-in, controllable potential (although the cantilever torsion might provide one). The readout is indirect.

\subsection{The quantum-opto-mechanics experiments}

The exemplar is the “micromechanical oscillator”, a small beam or bridge clamped at both ends and free to vibrate.
 Markus Arndt, Markus Aspelmeyer, and Anton Zeilinger wrote in 2009, \cite{Arndt} (see also \cite{ASP}):

\begin{quote}
The developing field of quantum-opto-mechanics provides ... a unique opportunity to generate superposition states of massive mechanical systems ... 
one arrives at the canonical situation of Schrödinger's cat involving two macroscopically distinct motional states of a mechanical resonator.
\end{quote}

The prefix ``opto-"  references optical; 
as for the Marshall \ea\ experiment, 
they envision photons reflected off the beam to make the cat.  
However, there remains ``the intriguing question whether it will be possible to generate macroscopic displacements 
that exceed the physical size of the mechanical object". It will be soon for ``carbon nanotubes or a silicon nanowire", which might restrict to the quantum side of the IB. 
However, the Viennese physicists remark that the imagined machinery spans the size range

\begin{quote}
from hundreds of nanometers ... to tens of centimeters in the case of gravitational wave antennae. It is currently a hot research topic how to prepare genuine quantum states of motion of such mechanical devices.
\end{quote}

Thus with these proposed ``quantum machines" scalability is available, but there remains the displacement issue, 
and we have to ask for the details of the amplification step and whether it is controllable.

\subsection{The Abdi \ea experiment}

In 2016, a German-British collaboration proposed, \cite{abdi}, to conduct an experiment on a ``lithium-decorated monolayer graphene sheet" of diameter one micrometer
suspended in a ``controllable, electrostatic double-well potential"”. The metallic lithium will render the wafer electrically conductive. 
The authors propose to cool the system to milliKelvin temperatures, aiming to attain the ground state. 
Observation is by way of magnetic coupling to a ``superconducting qubit", with which they hope to explore ``higher-occupation number states" (Copenhagenist language; for Schr{\"o}dingerists,
higher-frequency modes). The goal is to test conventional linear QM vs. QM + CC.

One micron is smaller than a human cell, but larger than a virus. It may lie on the classical side of the IB. Scalability is an issue. 

\section{Discussion}

The experiments summarized above have size and displacement limitations 
(for a review of the ``macroscopicity" achievable by various other suggested or conducted experiments, see \cite{arndthornberger}), 
and interpretational problems.
As the Abdi \ea experiment approaches most closely the ``ideal" case for purposes of testing the Hamiltonian theory against rivals, I discuss it in more detail here.

Leaving aside the scalability issue, the principle difficulty is the indirect observation of dispersion. 
Presumably the presence or absence of a cat will have to be deduced from computations combining many observed modes (known as ``quantum state reconstruction"). 
This generates an epistemological dilemma (which is not limited to this particular experiment).

Let us suppose an anomaly emerges from the experiment. In this context, this would mean a divergence of some measured quantities, here oscillation modes
 (those ``higher-number occupation states") from those predicted, say as a function of the ``anharmonicity parameter" (which I labelled ``height" in previous sections). 
The question is what to make of it.
Does it mean that QM, or QM+CC, is falsified? Let me sketch briefly the steps in the computation Abdi \ea list in an appendix to their paper:

\begin{quote}

(1) The quantum (\Schs) equation is replaced by a master equation.

(2) The equation is expanded in a Dyson series, retaining terms up to second order (Born approximation).

(3) Adopting an ```adequate' microsopic model for the system-environment interaction", 
 environmental states are introduced but reduced by ``truncating at a certain  number of states which are required in order for our simulations to converge".

 (4) Assuming the interaction with the environment is small, more terms are dropped (``rotating wave approximation").

 (5) Making more assumptions about memory in the environment, another, Markov, approximation is introduced.

 (6) Imaginary terms are dropped and a time integral extended to infinity.

 (7) Finally, the Markov model is simulated numerically.

\end{quote}

The dilemma must now be clear to the reader: is the outcome really anomalous? Or is one of the approximation methods inadequate, or an assumption about an ``environment" incorrect?
(Of course, computational ambiguity arises with any multi-body quantum system, 
since we cannot solve \Schs\ equation exactly except for systems with few degrees of freedom, or even simulate effectively with today's supercomputers. As for the ``environment": see next paragraph.) 
I believe there would be sufficient doubt that no one would abandon their paradigm
from such evidence alone. 
The cure is to introduce some direct readout of ``cat-or-no-cat", say by lowering in a microscope equipped with a weak light source and taking a snapshot of the wafer.
I am aware that this intervention would likely heat up the system, but perhaps the picture could be taken at the end of each ``run". 

I have not discussed ``decoherence" due to a putative ``environment", as I do not accept this theory as a solution of the Measurement Problem. Eliminating macroscopic interference
is the job of the apparatus (by cleanly separating wavepackets), not a mysterious ``environment".
Nor does replacing a superposition by a mixture supply a definite outcome---that's imposed by the nonlinear terms in the Hamiltonian, which force the system to make a choice, see paper III. 
If ``decoherence" exists I would regard it as a nuisance of the type that afflicts all experiments, i.e., external noise. 
I leave suppressing all such environmental perturbations to the skill of the experimentalist.

Even if an anomaly in the direction predicted by the Hamiltonian nonlinear theory (e.g., Figure \ref{dispfig}) appeared in the data, it would not ``prove" that theory true. 
As Karl Popper pointed out in 1935, \cite{popper}, 
data can falsify theories but never validate them. (Thomas Kuhn expressed doubt even about the falsification claim, \cite{kuhn}.) I would accept a failure to observe anything like the 
``hockey-stick graph" near the Infamous Boundary as a falsification of nonlinear QM.

Even to be relevant to the debate, data must be sufficiently ``clean" to permit unambiguous interpretation, and theory sufficiently rigid not to allow wiggle room for supporters to dismiss an anomaly.
I do believe that experimentalists are getting closer to performing an informative experiment.

\section{Computational Appendix}

The parameterization used for the potential was the following:

\def\hei{\hbox{height}}
\bar
\no V(x) &\=& A\,(x - \hbox{width})^2\,(x + \hbox{width})^2,\\
A &\=& \hbox{\hei}/\hbox{width}^4.
\ear 

Parameter "width" was 4.0. Other parameters were as in paper III. 
Figure \ref{dispfig} displays the peak dispersion over the time interval $[1,10]$ (arbitrary time units); the first interval was ignored because in the nonlinear case
the dispersion dropped from that of the initial condition. (``Dispersion" means of the ``center-of-total-spin" and included the square-root, like a standard-error 
rather than a variance, although for \Schists\ it isn't either.)

I also made simulations with a two-part function, $V_1(x)$, of form (writing $V(\hei,x)$ for the above):

\bar
\no V_1(x) &\=& V(10,x), \ph \hbox{for $|x| > 4.0$};\\
\no V_1(x) &\=& V(\hei,x), \ph \hbox{for $|x| <= 4.0$}.
\ear

I also tried diminishing the coupling constant of the ``micro" system to the ``macro", $\alpha$ in previous papers, by a factor of ten. Neither change altered Figure \ref{dispfig} substantially.

\begin{figure}
\rotatebox{0}{\resizebox{5in}{5in}{\includegraphics{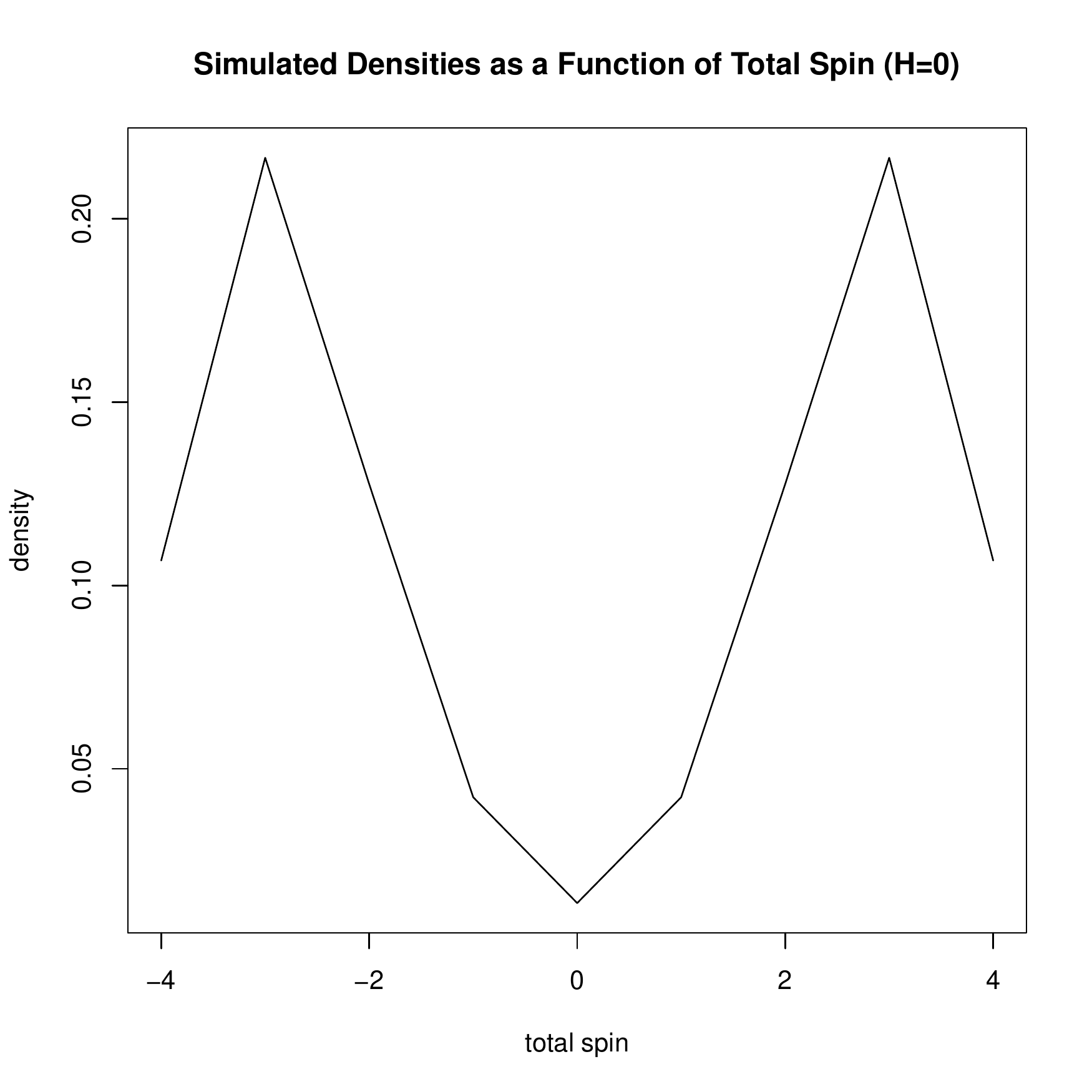}}}
\caption{Density vs. total spin for the linear case with zero external potential (seen at ``time 5", approximately the maximum dispersion).}\label{densfig}
\end{figure}

\begin{figure}
\rotatebox{0}{\resizebox{5in}{5in}{\includegraphics{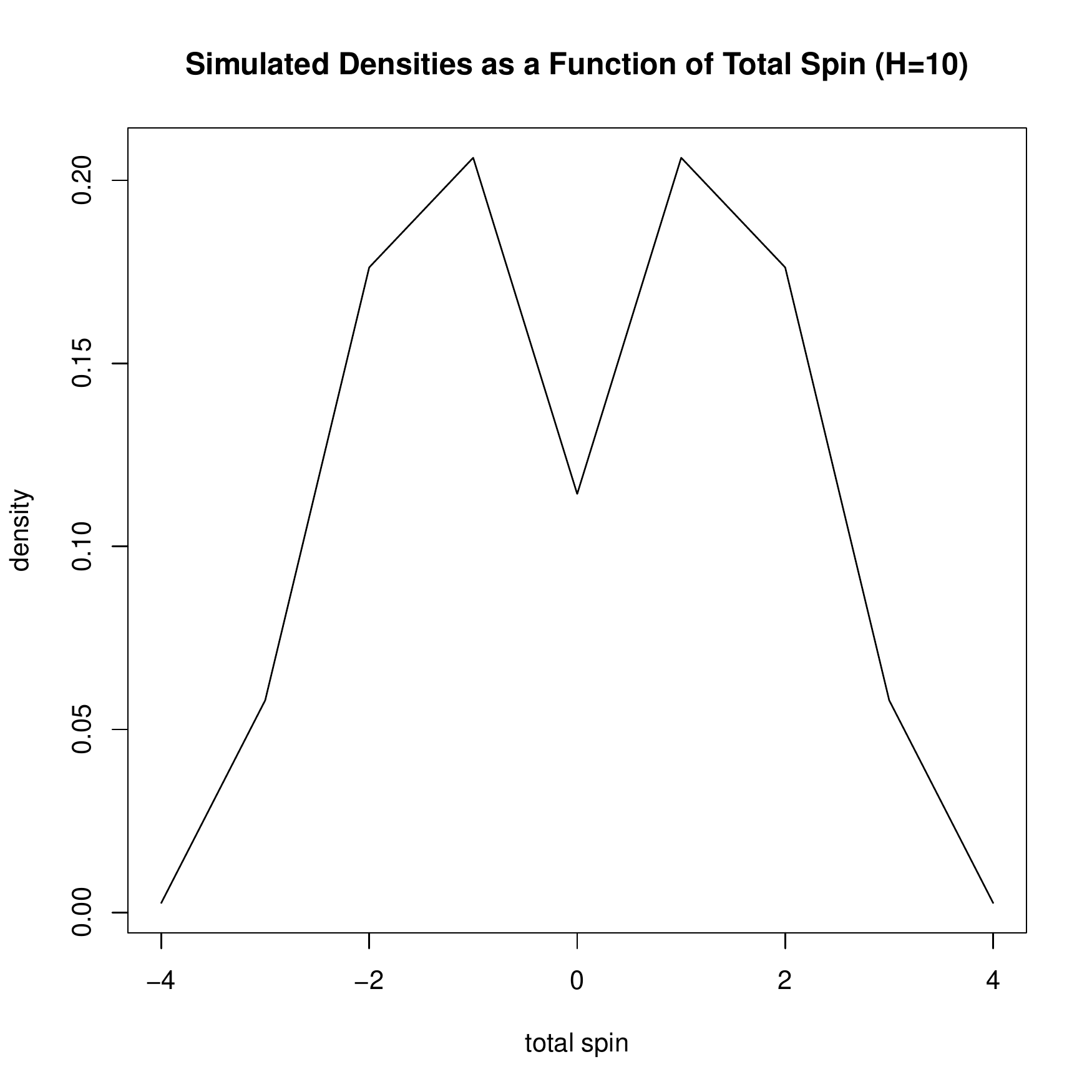}}}
\caption{Density vs. total spin for the linear case with external potential (seen at ``time 1", approximately the maximum dispersion).}\label{densfig2}
\end{figure}

The explanation for the trend of the dispersion in the linear case is that the double-well potential is not absolutely necessary to form cats, 
as the superposition of the up- and down-states in the ``microsystem" plus the linear coupling 
 to the ``macrosystem" can do the job.
The effect of adding the external potential is to partially confine the cats to the wells, actually diminishing the overall dispersion, 
see Figures \ref{densfig} and \ref{densfig2}. 
I suspect it is an artifact of the tiny size of the ``device" (8 qubits), and would not be reproduced at larger scales.  

Simulation used the Tao symplectic solver, described in the Appendix to paper III. 
Figure \ref{dispfig} required 6 hours and forty minutes running on a ten-year-old HP linux box. The program was written in the C language in the style of the old classic, 
{\em Numerical Recipes in C}; the reader is invited to reproduce (and extend) it on a modern platform with modern programming techniques.

\end{document}